**Quantum Features of Vacuum Flux Impact: An Interpretation of Quantum Phenomena**

**C. L. Herzenberg**

-------------------------------------------------------------------------------------------------

**Abstract**
Special relativity combined with the stochastic vacuum flux impact model lead to an explicit interpretation of many of the phenomena of elementary quantum mechanics. We examine characteristics of a repetitively impacted submicroscopic particle in conjunction with examination of the ways in which effects associated with the particle's behavior appear in moving frames of reference. As seen from relatively moving frames of reference, the time and location of impacts and recoils automatically exhibit wave behavior. This model leads to free particle waves with frequencies proportional to the energy and wavelengths inversely proportional to the momentum. As seen from relatively moving frames of reference, impacts and their associated recoils can appear to an observer to take place simultaneously at multiple locations in space. For superposed free particle waves corresponding to bidirectional motion, an amplitude that varies sinusoidally with distance results. A governing equation identical in form to the Schroedinger equation is developed that describes the behavior of the impacts and their associated recoils. This approach permits many features of quantum mechanics to be examined within an intuitively visualizable framework.

-------------------------------------------------------------------------------------------------

# Quantum Features of Vacuum Flux Impact: An Interpretation of Quantum Phenomena

C. L. Herzenberg

---

**Abstract**
Special relativity combined with the stochastic vacuum flux impact model lead to an explicit interpretation of many of the phenomena of elementary quantum mechanics. We examine characteristics of a repetitively impacted submicroscopic particle in conjunction with examination of the ways in which effects associated with the particle's behavior appear in moving frames of reference. As seen from relatively moving frames of reference, the time and location of impacts and recoils automatically exhibit wave behavior. This model leads to free particle waves with frequencies proportional to the energy and wavelengths inversely proportional to the momentum. For superposed free particle waves corresponding to bidirectional motion, an amplitude that varies sinusoidally with distance results. As seen from relatively moving frames of reference, impacts and their associated recoils can appear to an observer to take place simultaneously at multiple locations in space. A governing equation identical in form to the Schroedinger equation is developed that describes the behavior of the impacts and their associated recoils. This approach permits many features of quantum mechanics to be examined within an intuitively visualizable framework.

---

## 1. INTRODUCTION

Modeling a physical vacuum to include an intense flux of high-speed, electrically neutral, long-lived particles permeating space provides a mechanism that can reproduce a range of physical phenomena.[1] Previous work has shown that vacuum flux impact can account for features of classical gravitation and may be implicated in other phenomena such as dark energy. The present paper extends this work to examine in more detail some features of a vacuum flux impact model that lead to phenomena similar to those of quantum mechanics. While other studies have shown that combining classical mechanics with the stochastic mechanics of Brownian motion can formulate nearly all of the results of quantum mechanics,[2,3] the present paper presents a particularly visualizeable approach and emphasizes the role of particle impacts and recoils and the role of special relativity in the interpretation of quantum phenomena.

It has been found that the temporal quasiperiodic behavior of repetitive impacts of the vacuum flux on a matter particle leads to a statistically well-defined average frequency proportional to the particle mass.[1] For the case of a particle in uniform motion, special relativity leads to spatial as well as temporal periodicity, resulting in wave motion with parameters and characteristics similar to those of free particle quantum wave functions. We will examine how the Lorentz transformation effectively maps properties of the impacts and recoils along the direction of relative motion, irrespective of the actual spatial direction of the original impact or recoil in the particle's frame of reference. Because in the observer's frame of reference different impacts can present themselves simultaneously at different locations, a single particle can appear to manifest itself in many locations at once in this model. For bidirectional constant speed motion, sinusoidal state functions result. Combining an equation based on the rate of incidence of vacuum impacts on a particle with the general equation describing wave motion of a specific velocity leads to a governing equation with the form of the Schroedinger equation. Thus, special relativity together with stochastic vacuum impact provide for an explicit interpretation of some of the characteristic phenomena of elementary quantum mechanics.

## 2. BACKGROUND

The vacuum flux impact model proposes a specific physical cause or origin of stochasticity in stochastic quantum mechanics. In this model, we treat the physical vacuum as a real physical medium having as a component a high-speed, uncharged particulate flux that is pervasive in empty space. These particulates are characterized as collisionless with respect to each other; thus, in the absence of ordinary matter, these vacuum particulates travel undeflected through empty space. While these particulates do not interact with each other appreciably, they do interact with ordinary matter, primarily by the exchange of momentum in collisions. In this model, every material object is in continuing interaction with the vacuum particulate flux. The property of matter that we designate as mass turns out to be a measure of the interaction cross-section of ordinary matter with the vacuum particulate flux; thus, the vacuum particulate flux may be considered the origin of mass through its coupling with ordinary matter.[1]

Where matter is present, these vacuum particulates impact on particles of matter, causing recoils. Because the vacuum particulate flux is extremely high, the bombardment of ordinary matter particles is very intense, and the frequency of impacts is extraordinarily high. As a result, on an extremely small scale, the trajectory of an ordinary matter particle subjected to these impacts is very irregular and erratic. Continuing recoils manifest themselves as a zitterbewegung-like stochastic motion of individual matter particles. Thus, a randomness such as that characteristic of quantum phenomena results as a consequence of stochastic impacts from a component of the physical vacuum.

The vacuum particulate flux will impact on massive macroscopic objects through their constituent particles, imparting momentum impulses to them. A directional excess or deficit of vacuum flux impacting against a macroscopic object would cause acceleration in conformity with Newton's second law, as would any other source of momentum transfer. As a result of the interaction between matter and the vacuum flux, in the presence of matter the flux of vacuum particulates can become modified and exhibit anisotropy. The mutual shielding or shadowing of material objects from the vacuum flux by each other's presence leads to the occurrence of a flux-mediated inverse-square attractive force between them, so the principal features of classical gravitation result rather simply as features of the vacuum flux impact model. Thus, the interaction of ordinary matter with the vacuum particulate flux provides a simple and straightforward model for the origin of gravitational force.

Both gravitational forces and some quantum phenomena thus emerge from this model as medium-induced behavior, originating from the same medium but becoming manifest at very different scales of distance and time and particle number. The basic parameters characterizing this model turn out to be closely related to the Planck scale.[1]

## 3. VACUUM FLUX IMPACT EFFECTS ON AN OBJECT AT REST

The model is developed from an initial assumption that the flux of vacuum particulates is macroscopically omnidirectional, isotropic, and uniform as seen from the frame of reference of an isolated matter particle subject to vacuum flux impacts. The frequency of interactions of the flux with an individual particle of ordinary matter is evaluated in terms of an interaction cross section which is a measure of the probability of occurrence of a collision or interaction. The cross section is the reaction rate per unit incident flux of vacuum particulates per target particle of ordinary matter. Accordingly, the number of interactions per second of the flux with an individual object of ordinary matter will be given by:

$$f_i = \varphi \, \sigma \tag{1}$$

Here, φ is the flux (in vacuum particulates per square centimeter per second) and σ is the interaction cross section (in square centimeters). The flux φ is regarded as a constant quantity, except in the presence of gravitational potentials. The quantity σ is a basic characteristic property of the particle or object of ordinary matter subjected to bombardment by the flux of vacuum particulates. In previous work, in which the classical Newtonian law of gravitation was derived from the vacuum flux impact model, it has been shown that the cross section σ must be directly proportional to the particle's mass.[1]

Accordingly, the rate of occurrence of impacts or frequency of impacts on a particle will depend linearly upon the particle mass. Impact-related frequencies are found to be related to quantum frequencies associated with the particle mass, and are extremely high frequencies, even for the lowest mass particles; hence over macroscopic time intervals these processes reflect very large numbers of impacts and can be treated statistically. (Given how large the quantum frequencies associated with the rest masses of ordinary matter particles are, in accordance with the law of large numbers it would appear that significant fluctuations from average behavior must be rare.)

Since the frequency of impacts depends linearly on the particle's mass m, we can write:

$$f_i = m/b \qquad (2)$$

Here, b is a constant quantity that is expected to depend only on natural physical constants. Alternatively, using Einstein's mass-energy relationship $E = mc^2$, Eqn. (2) can be reexpressed as:

$$f_i = E/bc^2 \qquad (3)$$

(It may be noted that a specific relationship similar to Eqn. (2) or Eqn. (3) with a fully determined constant of proportionality could be arrived at directly by making some straightforward assumptions that we prefer not to make at this point in the analysis. The Planck equation $E = hf$ permits relating an energy to a frequency. Combining this with Einstein's mass-energy relationship leads to the de Broglie relationship that associates a frequency $f_m = mc^2/h$ with every particle of matter.[4,5,6] If we were to equate the characteristic vacuum interaction frequency to the de Broglie frequency, we would obtain an equation analogous to Eqn. (2) with a fully specified constant of proportionality. However, we wish to proceed independently of any such assumptions that might be regarded as containing any implicit quantum connections.)

As a result of Eqn. (2), the average time interval between impacts will be given by:

$$\tau_i = b/m \qquad (4)$$

The main results obtained in earlier work on the vacuum flux impact model are not dependent upon the details of the interaction of ordinary matter with vacuum flux particulates. However, some of the characteristics of these impacts may now be of further interest.

Because the vacuum flux particulates turn out to be extremely energetic and carry very high momenta on average, individual interactions are high-energy relativistic collisions and result in particle recoils that can occur at a speed close to the velocity of light. A particle of ordinary matter, in its stochastic motion, will have instantaneous speeds very close to the speed of light. Consequently, the distance traversed by the particle in an individual recoil corresponds on average closely to the product of the time between impacts and the speed of light:

$$d_i = cb/m \qquad (5)$$

Because the flux impacting a particle can come from all directions in three dimensional space, from geometrical considerations of solid angle it would appear that when a recoiling particle is struck by the next impact, it will on average tend to move in a direction perpendicular to its initial recoil direction. Thus, succeeding recoils will tend to be perpendicular to each other statistically.

Earlier results obtained from this model were confined largely to the case of stationary particles or macroscopic objects. [1] We now undertake more detailed examination emphasizing effects similar to quantum phenomena, beginning with examination of the case of a uniformly moving particle.

## 4. VACUUM FLUX IMPACT PHENOMENA FOR A PARTICLE IN UNIFORM MOTION

We associate a frame of reference with a particle of matter, such that the particle is at rest (in a time-averaged sense) with respect to its frame of reference. We synchronize clocks throughout this frame of reference. In this coordinate system, the particle is characterized by a specific frequency of impacts. How is this particle's behavior perceived from other relatively moving coordinate systems?

Let us examine this situation in which a matter particle is observed from a different frame of reference with respect to which it is in relative motion. Here, we are referring to time-average relative motion, corresponding to relative translational motion associated with two relatively moving coordinate systems, not to the instantaneous chaotic submicroscopic recoil motions associated with impacts.

In previous work on the vacuum flux impact model, we have relied on the validity of the special theory of relativity even in extreme circumstances, and we will continue to do so. For strict applicability of the special theory of relativity, we will in this section treat the average translational relative motion between the particle and the observer as constant and uniform for all time.

Here, we are not so much interested in the direct relationship between particle spatial coordinates in the two frames of reference, but rather in more global aspects of the relationship between times and time intervals in the two frames of reference. For the most part, we will disregard the actual detailed spatial locations of the particle as mapped onto the observer's frame of reference, and instead direct our attention to temporal events in the two frames of reference. We will emphasize examination of how temporal events associated with the particle would be observed in the laboratory frame of reference.

Initially, we will examine temporal synchronicity of impacts in moving frames. In the (time-average, not instantaneous) rest frame of the matter particle, the flux impacts occur at a statistically well-defined average frequency. We might regard this as defining a clock of sorts, which could provide a time standard throughout this frame of reference. As noted, we will consider clocks to be synchronized throughout this frame of reference. Thus, the times at which these impacts occur will be seen as simultaneous throughout a coordinate system associated with the time-average rest frame of the matter particle.

Next let's look at the continuing impact process as seen from another frame of reference with respect to which the matter particle is in relative translational motion. This frame of reference will be designated the observer's frame of reference or the laboratory frame of reference. As seen from the observer's frame of reference, there is no longer a simultaneity of these impact times throughout the observer's space. Rather, different impact times will be associated with different spatial locations in the observer's coordinate system in accordance with the Lorentz transformation equations.

Let us designate the (time average) location of the particle as specifying the origin of its coordinate system, and the location of the observer as specifying the origin of the observer's coordinate system. (As will be seen, the particular positions of the particle and the observer turn out in many respects to be largely irrelevant to the overall results; this will be discussed subsequently in further detail.) If the primed coordinate system associated with the matter particle is moving to the right with a constant uniform velocity V along the x axis as seen in the unprimed coordinate system associated with the observer, then, with the usual convention that the origins of the coordinate systems pass each other at mutual time zero, the Lorentz transformations give us: [7]

$$t' = \gamma (t - Vx/c^2) \qquad (6)$$

and:

$$t = \gamma (t' + Vx'/c^2) \qquad (7)$$

Here, as usual, the relativistic correction factor $\gamma = (1 - V^2/c^2)^{-1/2}$. For the time being, we will limit our attention to non-relativistic velocities; thus we can approximate adequately by setting the factor $\gamma$ equal to 1 and simplify the equations.

We note that the Lorentz transformation amounts to a mapping that can transform a sequence of temporal events in one frame of reference into a sequence of spatiotemporal events in another frame of reference. More specifically, a Lorentz transformation can map a sequence of temporal events in one frame of reference into a sequence of spatial events in another frame of reference at a specific time in the latter frame of reference.

We can see from Eqn. (7) that for a given value of the time (t) throughout the observer's frame of reference, increasing values of x' are associated with decreasing values of t' in the particle's frame of reference. Thus, the forward direction of the particle's frame of reference can be associated with its past and the backward direction with its future. Accordingly, at any moment of time (t) in the observer's frame of reference, in the particle frame the past is ahead along the direction of its motion relative to the observer's frame as seen from the observer's frame, while the future is behind, opposite to the direction of its motion. The appearance of the 'next' impact point is just behind, while the 'previous' impact point is just ahead along the direction of motion as seen from the observer's frame of reference.

**Spatial periodicity in the observer's frame of reference**

Each individual impact that occurs at a particular time in the primed frame of reference of the particle will be seen to occur at different times at different spatial locations in the observer's unprimed laboratory frame of reference. As a particular case, suppose that a vacuum flux impact occurs to the particle in the primed frame of reference at time t' = 0. This particular impact will be seen to occur at various different values of the time when observed from different spatial locations in the observer's frame of reference. In accordance with Eqn. (6), the times and locations in the observer's coordinate system associated with this event will be related by $t = Vx/c^2$. Thus, this particular event will be seen at earlier times at smaller values of x and at later times at larger values of x. (In effect, this 'event' will appear in the observer's frame of reference to move along the direction of relative motion at a speed $c^2/V$.)

Furthermore, the next impact, which will occur at the time in the particle's frame of reference of $t' = \tau_i$ will be seen in the observer's coordinate system at times and locations given in this non-relativistic approximation by:

$$t = Vx/c^2 + \tau_i \qquad (8)$$

where $\tau_i$ would be given by Eqn. (4).
Alternatively, we can reexpress Eqn. (8) as:

$$x = (t - \tau_i)c^2/V \qquad (9)$$

More generally, these periodic impacts occurring at particular times in the particle's frame of reference will be seen to occur in the observer's frame of reference at times and locations:

$$t = Vx/c^2 + n\tau_i \qquad (10)$$

or at spatial locations and times:

$$x = (t - n\tau_i)c^2/V \qquad (11)$$

in which n can take on all positive and negative integral values. Thus, at the t = 0, the periodic impacts would appear to take place in the observer's frame of reference at the locations:

$$x = -n\tau_i c^2/V = -n(bc^2/mV) \qquad (12)$$

and thus they can be seen to occur at locations corresponding to integral multiples of a specific distance interval that depends inversely on the momentum of the particle. We can identify the spatial repetition distance as:

$$L = \tau_i c^2/V = d_i(c/V) = bc^2/mV \qquad (13)$$

The resemblance to a de Broglie wavelength is evident.

What amounts to a wave number can be introduced as the inverse of this distance, as:

$$k_i = 1/L = mV/bc^2 \qquad (14)$$

This provides a spatial characterization of the periodicity as seen in the observer's laboratory frame of reference.

More generally, let us consider a periodic sequence of temporal events ("ticks") in a moving frame of reference, specified by $t' = 0, \tau_i, 2\tau_i, 3\tau_i, \ldots$. Let us consider mapping these with a Lorentz transformation onto an observer's coordinate system. Using Eqn. (6), they will be mapped onto $t = Vx/c^2$, $t = \tau_i/\gamma + Vx/c^2$, $t = 2\tau_i/\gamma + Vx/c^2$, $t = 3\tau_i/\gamma + Vx/c^2$, …. Thus, individual "ticks" in the particle's frame of reference map into specific moving events or signals in the laboratory frame of reference, all periodically separated from eachother and moving together at the same speed $c^2/V$, and each separated from its neighbors by spatial intervals $c^2\tau_i/\gamma V$. Thus, we can anticipate from the Lorentz transformation Eqns. (6) and (7) that if temporally periodic events occur in one frame of reference, that they will map to a spatially periodic moving distribution in another relatively moving frame of reference.

Accordingly, in the observer's frame of reference, a particular impact will be presented at different locations in space at different times. In addition, at each location in the observer's space, impacts will occur (and have occurred) periodically throughout time. Furthermore, the entire sequence of temporal intervals that occurs in the moving frame of reference will be mapped as spatial intervals at any given time in the observer's frame of reference.

Generally speaking, what we are seeing is that because any particle in the vacuum flux impact model is characterized by periodic impacts whose frequency depends on the particle's mass, the temporal behavior of such a particle is periodic with a well characterized period. Furthermore, as seen from a relatively moving frame of reference, the spatial behavior will also be periodically repetitive. Thus, periodic behavior is defined throughout all of space and time for this particle, as seen from a relatively moving frame of reference.

We may note that for a particle traveling in the opposite direction (toward – x values) at speed V or velocity –V, a similar calculation indicates that periodic impacts occurring at particular times in the particle's frame of reference will be seen to occur in the observer's frame of reference at similar times and locations characterized by the same expressions with a –V replacing V. These periodic impacts will be seen to occur at a corresponding set of locations also separated by the same spatial repetition distance L.

**Wave-like behavior of the relativistic representation of repetitive impacts in the observer's space**

Let us direct our attention to a particular impact that occurs at a specific time (e.g. $t' = 0$) in the particle's frame of reference. In the observer's frame of reference, such an event will be seen at earlier times for smaller values of x and at later times for larger values of x. In other words, for each impact that occurs in the particle's frame of reference, its mapping in the observer's frame of reference moves steadily forward along the x axis in unison with all other impact events. We might interpret this by saying that a discrete event in the particle frame of reference becomes a continuing phenomenon as seen in the observer's frame of reference. As seen from the observer's system, as time increases, this particular impact appears at larger values of x. Its manifestation in the observer's frame of reference occurs sequentially in time at sequential spatial locations, and thus this spatial impact pattern can be regarded as moving, as seen from the observer's frame of reference. As noted earlier, a particular impact will appear to move at a speed $c^2/V$ in the observer's frame of reference.

Why is this velocity greater than the velocity of light? That occurs because in the special relativistic mapping, there would be (c/V) actual recoil steps that occurred in the particle's frame of reference during the time interval associated with the Lorentz mapping of a single recoil interval onto the axis of relative motion in the observer's frame of reference. While the particle traverses an actual distance of approximately $c\tau_i$ in its frame of reference (recoiling in whatever direction is determined by the vacuum flux impact), the corresponding mapped event appears as simultaneously translated by a distance $V\tau_i$ along the direction of relative motion, as seen in the observer's frame of reference. The mapping gives a length (corresponding to a mapped image of a recoil) in distance units of $L = bc^2/mV = (b/m)(c^2/V)$, or $(bc/m)(c/V)$. This would correspond to (c/V) steps of length $d_i = c\tau_i$. Thus, the mapped image associated with a single impact and its recoil extends over a distance c/V times as long as the physical distance interval of the event that is mapped.

So what is the physical origin of the velocity $c^2/V$ that characterizes these waves? In the vacuum flux impact model, this velocity originates directly from the Lorentz transformations; - just as the relative velocity V is the coefficient of the second term in the equation for the transformation of spatial coordinates, so also the quantity $c^2/V$ is the coefficient of the second term in the transformation of temporal coordinates, and it is the mapping of the temporal features of repetitive impacts that leads to this and certain other distinctive features of quantum mechanics.

Next, let us consider the feasibility of introducing a function of space and time to describe the somewhat wave-like behavior of the relativistic representation of a free particle subject to vacuum flux impacts. Then, considering that this functional behavior originates from the periodic behavior of the particle as a function of its intrinsic time t', we could write such a function in the form $\varphi_1(t')$. Using Eqn. (6), we can reexpress that as a function of t and x as $\varphi_1(\gamma t - \gamma Vx/c^2)$, and in the non-relativistic approximation, this would become $\varphi_1(t - Vx/c^2)$. If we rewrite the functional dependence including the frequency as a factor so that the dependence will be expressed in terms of a product of frequency and time, this would become $\varphi_2(f_i t - f_i Vx/c^2)$, where $\varphi_2$ is a separate function very simply related to the function $\varphi_1$. Using Eqn. (2) and Eqn. (14), this becomes $\varphi_2(f_i t - k_i x)$, where $k_i$ is the associated wave number. Not surprisingly, we find that we are describing a functional similarity to a running wave, with parameters characterizing the periodicity of impacts. In dealing with cyclical phenomena, a phase is commonly introduced. The quantity ($f_i t - k_i x$) will then constitute an angular phase in radians as a function of x and t, that is descriptive of this periodic wave-like behavior.

Thus, for free particle motion, the vacuum flux impact model leads quite directly to the existence of what amount to waves exhibiting frequencies and wavelengths resembling those associated with the corresponding free particle wave function solutions of the Schroedinger equation.

**Where do we see the particle?**

In the vacuum flux impact model, every particle is characterized by its relationship with the vacuum flux, and in particular, by its interaction cross-section with the vacuum flux given by Eqn. (1). A particle's detailed behavior is determined by the impacts of the vacuum flux on it that cause recoil motion.

At particular times in the observer's frame of reference, impacts will appear to occur simultaneously at periodically spaced locations all along the direction of relative motion. These impact events apparently occurring simultaneously at different locations in the observer's frame of reference actually correspond to different impacts in the particle's frame of reference. But, from the point of view of the observer, it would appear as if the particle were being impacted in multiple locations simultaneously. Thus, the feature of the strangeness of quantum mechanics in which a particle can appear to be at multiple locations at once can actually arise simply as an artifact of special relativity in conjunction with periodically repetitive impacts occurring in the particle's frame of reference.

Thus, when we look at a particle subject to periodic impacts from a relatively moving frame of reference, in accordance with special relativity we find that impacts will appear to occur at multiple locations in space as well as periodically in time. To the extent that we interpret the presence of impacts at a particular position and time as indicating the presence of the particle, to that extent, the particle will appear to the observer to be located at multiple positions in space.

**Characteristics of inter-impact intervals in the observer's frame of reference**

To more fully characterize where the particle appears, we need to be able to describe the occurrences of impacts in space and time.

What we are looking at in the special relativistic mapping onto the x-axis for cases in which relative motion exists between the particle rest frame and the observer is a mapping onto the x-axis of the periodic impacts and times associated with recoils between them. The time intervals between impacts in the particle's frame of reference correspond to time intervals associated with the particle recoils. Accordingly, we may regard these recoil times as also mapped onto the x-axis between succeeding impacts.

During actual recoil time intervals, two distinct types of motion are present: the translational motion associated with the relative velocity of the frames of reference, and the submicroscopic recoil motion. The translational motion is uniform, occurring along the axis of relative motion of the separate frames of reference, while the particle recoil motion occurs stochastically in various directions in 3-dimensional space. We will examine how both types of motion might be represented in the time intervals between mapped impacts.

The translational motion in association with the periodic phenomenon can be represented directly using the special relativistic transformation and mapping that we have been examining already. The question then arises how actual particle recoil motions might be mapped in association with the temporal relativistic mapping. The mapping associated with each impact might be expected to appear in the spatial intervals between impacts in the observer's frame.

Temporally sequential impacts in the particle's frame of reference also appear spatially sequential in the observer's frame of reference. Thus, a series of temporally sequential impacts is mapped sequentially along the x axis. While in the particle's rest frame the impacts may occur along various different directions in three-dimensional space, their representations as seen in the observer's frame of reference are all aligned along the x-axis.

Because special relativity 'rotates' between t and x when relative motion occurs along the x-axis, it maps time intervals associated with a particle's motion onto spatial intervals in the observer's frame of reference. Thus the time intervals associated with erratic motion (e.g. a random walk) would, as seen in a relatively moving frame, be mapped along the x-axis, irrespective of their direction in the particle's rest frame. A Lorentz transformation could effectively map the time interval associated with each impact onto the observer's x-axis (by virtue of the relationship joining x and t), irrespective of the actual spatial direction of the original impact or recoil in the particle's frame of reference. The jagged path or random walk performed by the particle within its rest frame under the influence of repeated impacts would not be reproduced directly, rather, the time of each impact would be mapped by the Lorentz transformation into a straight line along the x axis in the observer's space. Indirectly, the lengths of recoils (through their associated time intervals) could be mapped onto the observer's x-axis and show up in terms of the wavelength of the periodicity.

## 5. VACUUM FLUX IMPACT PHENOMENA FOR A PARTICLE IN BIDIRECTIONAL MOTION

One of the simplest physical systems analyzed in elementary quantum mechanics is the case of a particle moving to- and fro- along one dimension at constant speed. We can examine this system of bidirectional motion using our modeling approach.

Let us consider a situation in which uniform motion takes place at the same speed in opposite directions simultaneously. In attempting to understand the system, it is useful to examine both the separate motions and the different possibilities of combined motions.

This case of to- and fro- motion can be at least partially examined with the help of special relativity by considering the two states of motion separately. While special relativity is strictly speaking applicable only to constant velocity motion, we can still use special relativity as a tool to explore aspects of other types of motion so as to acquire insight and perspective, as long as we bear in mind the limitations of this approach.

The issue as to how these two states of motion normally associated with separate identical particles will eventually be regarded as pertaining to a single particle will be deferred for now. We can implement this approach by examining possible ways of combining particle motions representing both translational motions and submicroscopic recoil motions.

Let us initially consider what we may be able to learn about to-and fro- translational motion by examining relativistic results for uniform translational motion in each direction. We can use the Lorentz transformation expressed in Eqn. (6) to evaluate the intrinsic time variables in each of the two moving frames of reference, and to express them in terms of the observer's (or laboratory) coordinates. The time $t_+$ in a system moving to the right with respect to the laboratory frame of reference at speed V is related to the time t and location x in the laboratory frame of reference by the equation:

$$t_+ = \gamma(t - Vx/c^2) \tag{15}$$

The time $t_-$ in the system moving to the left with the same speed with respect to the laboratory frame of reference is related to the time t and location x in the laboratory frame of reference by the equation:

$$t_- = \gamma(t + Vx/c^2) \tag{16}$$

We see that, although time passes at the same rate in the two moving frames of reference, the actual values of the time parameters differ from each other at any particular time and spatial location in the laboratory (observer's) frame of reference.  Thus, for to- and fro- motion, it would appear that no common time could be established throughout the observer's space on the basis of using special relativity.

However, let us consider forming an average time by equally weighting and combining the time parameters associated with the forward and the backward translational motions in accordance with special relativity.  Combining the preceding equations, we would find for such an average time associated with to- and fro- motion:

$$t_{avg} = (t_+ + t_-)/2 = [\gamma(t - Vx/c^2) + \gamma(t + Vx/c^2)]/2 = \gamma t \qquad (17)$$

Thus, we see that this average time associated with to- and fro- motion is the same everywhere throughout space.  Furthermore, it is equal to laboratory time as modified by the relativistic correction factor $\gamma$.  Since this relativistic correction factor approaches a value of one for the case of non-relativistic velocities, the average time associated with to- and fro- motion approaches the laboratory time in the non-relativistic limit.  Such an average time associated with to- and fro- motion could be interpreted in the limit as the time corresponding to a particle at rest.

**Impacts, spatial periodicities and wave-like behavior associated with bidirectional motion**

In order not to lose any part of the information contained in the two separate time parameters $t_+$ and $t_-$ associated with the opposite directions of translational motion, we may not only employ the average time parameter, which is based on the sum of the two individual time parameters, but also take into consideration the difference between the two individual time parameters.

We can introduce a parameter based on the difference between the two separate time parameters associated with the to- and fro- motions and evaluate it using Eqn. (15) and Eqn. (16):

$$t_{diff2} = (t_+ - t_-)/2 = [\gamma(t - Vx/c^2) - \gamma(t + Vx/c^2)]/2 = -\gamma Vx/c^2 \qquad (18)$$

We see that this parameter based on the difference between the individual time values for to- and fro- motions actually turns out to be independent of the time and exhibits only a dependence on the spatial variable x in the laboratory frame of reference. This parameter may be considered in addition to the average time, because taken together these parameters based on the sum and difference retain the full original information in the two separate time parameters $t_+$ and $t_-$.  We see that forming the parameters related to the sum and difference of the time parameters associated with to- motion and the fro- motion, allows us to separate an average temporal behavior from a spatial behavior in the laboratory frame of reference.

We can anticipate that there may be manifestations of spatially periodic behavior in bidirectional motion because there are common spatial periodicities associated with the separate forward and backward states of motion that combine to form bidirectional motion.  We have already seen that the to- and fro- motions separately exhibit the same periodicities along the x-axis, with inter-impact intervals of equal length, given by $L = dc/V$ in the non-relativistic limit, corresponding to the mapping of a single recoil distance.

Impacts occur in the moving frames of reference at particular values of the times $t_+$ and $t_-$, e.g. at times 0, $\tau_i$, $2\tau_i$, etc.  In the non-relativistic approximation, at these times impacts would be seen to take place simultaneously at arrays of periodic locations along the x-axis in the observer's laboratory coordinate system.  Also, impacts would take place at these same values of average time at these same arrays of locations, in the non-relativistic approximation.

We have seen that the relativistic representation of repetitive impact times occurring in a moving frame of reference exhibits a wave-like behavior in the observer's frame of reference, characterized by a spatial repetition distance $L = dc/V$ and an apparent propagation speed $c^2/V$.  When two frames of reference moving at the same speed in opposite directions are present simultaneously with periodic impacts occurring in each, then they would each separately exhibit such wave-like behavior that propagates at equal speed in opposite directions and exhibits equal intervals of spatial periodicity. Accordingly, at the times that impacts occur (in laboratory time or in the average time in the non-relativistic limit), the two moving wave-like event manifestations may be expected to be in agreement in regard to specification of the locations of a uniformly spaced array of impacts along the x-axis.

Also, comparing the parameter $t_{diff2}$ in Eqn. (18) with Eqn. (11) and Eqn. (12), we can see again that successive periodic impacts occurring at particular times in the particle's frames of reference, separated by an interval $\tau_i$, could be expected to be manifest in the spatial presentation along the x-axis as spatial impact locations separated by inter-impact intervals of length L.

The vacuum impacts on a particle can be expected to continue at an essentially constant rate indefinitely, irrespective of the nature of the macroscopic motion of the particle. In this physical situation, the particle would continue to undergo periodic impacts, and the overall frequency at which impacts occur must remain the same independent of the directions of motion or how the motions are combined. Specifically, if the particle "doubles back" on itself in coordinated macroscopic to-and-fro motion, the overall impacts must continue at the same rate. How would these impacts be allocated between the different states of motion?

If we combine unidirectional motion along each direction of the x-axis to model bidirectional motion, the associated sets of impact times and impact locations would necessarily be coordinated in some manner. It seems reasonable to assume that the impacts would occur with equal frequency in association with the two translational states of motion; that is, that they would be allocated equally between the forward and backward states of translational motion. One could imagine each impact being shared, which would seem to imply that these two states of motion would be synchronized in the sense that they would have impacts at the same time, and that these impacts might also map at the same locations along the x-axis. As an alternative or additional possibility, impacts might be allocated individually between the to- and fro- states of motion. The most equitable way of sharing impacts between these states would appear to be to have alternate impacts occur in association with the forward and the backward translational states of motion. In this case, the frequency of impacts associated with motion in either direction would be half the original impact frequency, and the impact locations would be separated by twice the distance, and alternate between each other. Thus, the mappings of two recoil lengths could be accommodated between the comparable impact locations. In this case, impacts would occur alternately at sets of points along the x-axis that correspond to the two meshed sets of impact points. That is, impacts would occur on one set, then at the next set, then on the first set of points again, and so on. This would appear to provide a manifestation of a kind of standing wave.

Earlier, we considered the possibility that the wave-like behavior of the relativistic representation of a particle subject to vacuum impacts moving at constant velocity might be described by a particular function of space and time, and found that its temporal and spatial aspects could be described by a function $\varphi_2(ft - k_i x)$ characteristic of a moving wave. If we were to consider an analogous argument using the average time parameter $t_{avg}$ in Eqn. (17) as an intrinsic time associated with the particle in bidirectional motion, we would find that the particle behavior might be described by a function of the laboratory time, in the form $\varphi_3(ft)$, while the associated spatial dependence of its behavior that appears in Eqn. (18) would have to be present as a spatial function, presumptively in the form of a product. Accordingly, we may anticipate dealing with functional characteristics of a standing (rather than running) wave in describing the behavior of a particle in bidirectional motion.

**Recoil motion in association with spatial mappings**

We have seen that the time intervals between periodic impacts can be expected to appear in Lorentz transformation onto the x-axis as spatial intervals between impact locations.

During the time between impacts, submicroscopic recoils can be regarded as taking place in association with each direction of macroscopic motion. The recoil times associated with oppositely directed translational motion would then be mapped along opposite directions along the x-axis. Furthermore, particle motion directed along the positive x-axis would have future impact locations appearing along the negative x-axis and be characterized by a wave moving toward positive values of x, while particle motion directed along the negative x-axis would have future impact locations appearing along the positive x-axis and can be characterized by a wave moving toward the negative x-axis.

It may be noted that when impacts alternate between the two macroscopic translational motions, so that the impact times associated with motion directed along the positive x-axis fall midway between the impact times associated with motion directed along the negative x-axis, then the corresponding inter-impact periods would be out of phase, so that inter-impact periods associated with the two directions would overlap in the middle. As a result, it would appear that the halves of each individual inter-impact period associated with one direction of macroscopic motion would be able to combine with halves of adjacent inter-impact periods associated with the other direction of motion, allocating time for combined recoils in association with each direction of translational motion.

In association with spatial intervals created by the Lorentz transformations along the x-axis there would exist submicroscopic recoils oriented along various directions in space (we may not even know in principle what those directions are). Thus, associated with an inter-impact interval along the x-axis at any instant of time, there would be two recoils, one associated with translational motion toward the right (directed along the + x direction) and the other associated with translational motion toward the left (directed along the – x direction). In other words, in association with each inter-impact distance interval, there would be two individual submicroscopic recoil motions associated with translational motion directed along the + x and – x directions respectively.

**Composite recoil formation and motion**
  But we are not actually dealing simply with two particles moving in opposite directions along the x-axis and undergoing separate recoils, but rather with a single particle performing a combined motion.

A superposition of states would seem to be a manifestation of a quantum mechanical state in which – to put it bluntly – the particle keeps trying to do two different things at once. And not only is it trying to undergo equal and opposite macroscopic translational motions, but also, at the submicroscopic level, it is trying to undergo two different recoil motions simultaneously.

Accordingly, we will want to model a particle undergoing bidirectional translational motion during which it will simultaneously have to undergo two submicroscopic recoil motions. What is the nature of the net motion of the particle?

We may expect that the net resultant composite motion of the particle will depend upon the relative orientation of the two component recoil motions. These two separate recoil motions would have on average the same duration and each would have a speed close to c, the velocity of light. We could anticipate that the two submicroscopic recoil motions would combine on the basis of their relative orientation, and the nature of the net resultant composite submicroscopic recoil motion would then be expected to depend on the angle between the two individual participating recoil motions.

How should we approach modeling this? We could try modeling the particle in a superposed state of motion as moving along the direction of a resultant of time-shared velocities in submicroscopic motion as well as in translational motion. In the submicroscopic recoil motion, the particle might be taken to move in a shared motion along a trajectory midway between the trajectories associated with the two individual component recoil motions. We could approach modeling this by treating the net composite submicroscopic motion as occurring along the direction of the resultant vector of the two individual component velocity vectors.

Let us designate the geometrical opening angle between the two component recoil velocity directions as $\theta_g$. We will also employ the half-opening angle, $\theta_h = \theta_g/2$. These angles provide a measure of the relative orientation or correlation or coordination between the separate recoil motions that combine to form a composite recoil. From symmetry considerations, the net composite velocity of the particle might be expected to lie in the same plane midway between the two component vectors. One of the component recoil velocity directions would be at an angle $\theta_h$ to the net composite recoil velocity vector, while the other component recoil velocity vector would then be directed along the angle $-\theta_h$ with respect to the net composite recoil velocity direction in their common plane. One of these component recoil velocities (directed at the angle $\theta_h$) would be associated with translational motion directed toward positive values of x along the x-axis, while the other component recoil velocity (directed at the angle $-\theta_h$) would be associated with translational motion directed toward negative values of x.

In addition to the direction of the composite recoil motion, we will also be concerned with the speed of the composite recoil motion. The simplest and most obvious assumption we can make is that the speed of the net composite motion will depend on the half-opening angle $\theta_h$ between the individual component motions. That is, $v = v(\theta_h)$. (We use a lower-case v to distinguish the submicroscopic particle recoil velocity from the translational velocity, designated by an upper-case V.) We may expect that there will be a unique relationship between the opening angle and the composite recoil velocity.

We may anticipate that the net composite recoils (that originate from combining the individual recoil motions associated with each direction of macroscopic motion) will in most cases be moving more slowly than individual recoils. If, for example, the two individual submicroscopic recoil motions were to take place along opposite directions (corresponding to $\theta_h = \pi/2$), then the resultant net composite recoil motion might be expected to be at zero net velocity, so that the particle would not undergo any recoil motion during the composite recoil time interval. On the other hand, if the two individual recoil motions were aligned (corresponding to $\theta_h = 0$), then the composite motion of the particle might be expected to align with the common direction, and occur at a recoil speed very close to c. For intermediate values of the angle between the two submicroscopic recoil motions, the net recoil motion might be expected to be along a direction midway between the two individual recoil directions, and to occur at speed less than c.

Note that the angular range for $\theta_h$ between 0 and $\pi/2$ radians corresponds to net composite recoil speeds between c and 0. The angular range of the relative orientation angle $\theta_h$ between 0 and $\pi$ radians would include two mappings of the recoil range.

It therefore seems reasonable to conclude that the maximum recoil distance will be associated with a value of $\theta_h$ of 0 and that the minimum recoil distance of zero will be associated with a value of $\theta_h$ of $\pi/2$, with intermediate angles corresponding to recoil distances of intermediate length. Thus, we can expect the recoil distance for a composite recoil to be correlated uniquely with the opening angle between the combining individual recoil motions. Similarly, we might expect mappings of recoil distances along the x-axis to be correlated uniquely with the opening angle between the combining individual simple recoil motions.

We have seen that Lorentz transformations associated with to- and fro- motions define intervals along the x-axis that are associated with the inter-impact times and consequently may also be considered in relationship with the associated recoils. Let us depart from our emphasis on relativistic time values in order to briefly examine a direct spatial representation of recoil velocities in a spatial mapping of recoil distances along the x-axis. In such a spatial mapping (not a Lorentz transformation), an actual recoil length would be mapped over a spatial distance extending from an impact position to a distance corresponding to the mapped value of $v\tau_i$ where $\tau_i$ is the time associated with an impact in the laboratory frame of reference in the non-relativistic limit. This would correspond to a distance along the x-axis that would be in the ratio of (v/c) to the full mapped recoil distance. This provides a one-to-one relationship between composite recoil velocities and distances within inter-impact intervals along the x-axis.

**Composite recoil velocity magnitudes and associated relative orientation angles**

We are examining what amounts to a quantum mechanical problem of combining states and attempting to deal with it by using a classical approach. How does the particle interact or correlate with itself under these circumstances? We don't know with assurance how velocities would combine in the formation of composite recoils, but we can explore some reasonable possibilities.

Perhaps the most straightforward and simplest hypothesis and quantitative assumption that could be made here would be that the net speed of the composite motion of the particle would be given by the average value of the time-shared contributing component velocities along the resultant direction, that is, (in the approximation that the individual recoil motion takes place at a speed close to c) by:

$$v = c \cos \theta_h \tag{19}$$

If we re-express Eqn. (19) in terms of the complementary angle to $\theta_h$ which we will designate $\theta_c$ (where $\theta_c = \pi/2 - \theta_h$) then we can see by expanding in a Taylor series that for small values of $\theta_c$ corresponding to small non-relativistic composite velocities, that the composite velocity would just be proportional to the angle:

$$v = c \cos \theta_h = c \sin \theta_c = c \, \theta_c + \text{higher order terms} \tag{20}$$

Thus, using this straightforward assumption, we find that at least for small non-relativistic values of the composite velocity, we might expect a proportionality between the velocity and the complementary angle. This could provide a basis for an initial hypothesis or assumption that the geometrical angle that emerges from recoil motion velocity combination could be put into correspondence with a phase angle associated with the time or distance that would appear in a function describing the wave features of this phenomenon.

**Durations of composite recoils moving at different velocities**

In looking at the case of a free particle moving at constant velocity, we found it informative to examine relativistic time intervals and the spatial intervals that are associated with these time intervals by Lorentz transformations.

Now, we will examine in more detail the role of time intervals associated with composite recoils. In order to analyze the problem, we will proceed by separating the overall particle velocity into two components: a non-relativistic translational motion and a recoil motion that may be relativistic.

We will approach the problem by using a time-averaged rest frame associated with the particle motion (either the laboratory frame of reference corresponding to zero net translational velocity, or a frame of reference in relative translational motion at speed V moving toward the right or toward the left). Then only the composite recoils that are formed from oppositely (or nearly oppositely) directed individual recoils will take place at rest (or nearly at rest) with respect to such coordinates. Other composite recoils will take place at higher velocities.

Because of the differing velocities associated with different composite recoils, it may be worthwhile to look at time intervals in different relatively moving frames of reference, and take into consideration relativistic time dilation, which can have an important role when time intervals are observed in frames of reference moving at different velocities.

In special relativity, every clock appears to go at its fastest rate when it is at rest relative to the observer; whereas if the clock moves relative to the observer with a velocity v, the rate appears slowed down by the factor $(1- v^2/c^2)^{1/2}$.[8] As a consequence of the Lorentz transformation, an observer will estimate that a clock moving relative to him/her goes slow with respect to his or her own clock. As an experimentally confirmed example of the fact that time intervals in rapidly moving frames are elongated in accordance with special relativity, unstable particles that are moving rapidly with respect to the laboratory are observed to exist for longer times (durations) before decaying than such particles do when at rest (or nearly so) with respect to the laboratory; the modification in duration as a function of particle speed is observed to fit the Lorentz factor, $\gamma = 1/(1 – v^2/c^2)^{1/2}$, as expected from special relativity; thus, time intervals associated with a particle traveling at higher speed are observed experimentally as lasting longer than time intervals associated with a particle traveling at lower speed.[9]

Different composite recoils will take place at different velocities with respect to a time-averaged particle frame of reference, ranging from zero relative velocity to velocities near the speed of light, depending upon how individual recoil motions combine to form a composite recoil motion. In accordance with special relativity we would expect that intervals of time during such composite recoils would be extended. The time duration for such a composite recoil traveling at speed v would relate to a corresponding time interval as seen in a time-averaged particle frame of reference by the relationship:

$$t_D(v) = t_{tapD} (1- v^2/c^2)^{-1/2} \qquad (21)$$

Here, $t_D(v)$ is the duration of a composite recoil traveling at velocity v, while $t_{tapD}$ is the duration or recoil time in the time averaged particle frame of reference. Thus a particle engaged in a zero velocity composite recoil and hence at rest in the time-averaged particle frame of reference would exhibit a recoil duration $t_{tapD}$. For a composite recoil traveling a higher speed with respect to the time averaged particle frame of reference, the recoil time duration clearly would be longer than $t_{tapD}$, in accordance with Eqn. (21).

We are limiting our consideration of translational velocities to those that are small compared to the velocity of light, and in this non-relativistic limit, the recoil time in any time-averaged particle frame of reference will be equal to or very closely approximated by the recoil time in the laboratory frame of reference. Accordingly, we can reexpress Eqn. (21) as:

$$t_D(v) = \tau_i(1- v^2/c^2)^{-1/2} \qquad (22)$$

**Average duration of composite recoils**

The preceding discussion applies to the time duration associated with a single individual composite recoil. However, because of the extremely high repetition rate at which impacts occur (of the order of $10^{20}$ times per second if the particle is an electron, for example), any actual macroscopic laboratory observation of a particle would necessarily be averaged over very large numbers of individual composite recoils traveling at different velocities.

We can form an average duration of composite recoils by integrating over various contributing recoil velocities. Let us designate such an average time duration by $t_a$. Then, assuming that all of these mappings of recoils would start at an impact position or impact time in calculating these superpositions, we find:

$$t_a(v) = (1/c) \int t_D(v) \, dv \qquad (23)$$

where the integration is performed from zero velocity to a velocity value v that constitutes the upper limit of integration. Thus:

$$t_a(v) = (1/c) \int \tau_i (1 - v^2/c^2)^{-1/2} \, dv \tag{24}$$

Here $t_a(v)$ is the average time duration associated with the velocity v that corresponds to the upper limit of integration.

If we evaluate this integral, we find: [10]

$$t_a(v) = \tau_i \arcsin(v/c) \tag{25}$$

or:

$$t_a(v) = -\tau_i \arccos(v/c) \tag{26}$$

But in accordance with Eqn. (19), $v/c = \cos\theta_h$. Consequently, in magnitude:

$$\arccos(v/c) = \theta_h \tag{27}$$

Here $\theta_h$ is the half-angle of separation between the two recoil velocity vectors contributing in combination to form a composite recoil. Also:

$$\arcsin(v/c) = \theta_c \tag{28}$$

where $\theta_c$ is the angle complementary to the half opening angle $\theta_h$ (that is, $\theta_c = \pi/2 - \theta_h$).

Combining Eqn. (25) with Eqn. (27), we find that we can express the average magnitude of the time interval during superposed composite recoils as simply proportional to the relative orientation angle $\theta_c$:

$$t_a(v_i)/\tau_i = \theta_c(v) \tag{29}$$

Here $\theta_c$ is a function of the value of the velocity v that is the limit of integration in Eqn. (24).

Eqn. (29) provides an evaluation of an average or apparent time interval for recoils, averaged over the contributing velocities for many composite recoil events. This result enables us to relate an average time after impact to an angle associated with the formation of composite recoils.

We see that the time interval during composite recoil, averaged over contributing velocities, would just be proportional to the relative orientation angle $\theta_c$, which is complementary to the half opening angle $\theta_h$ between the component velocity vectors. Note that an individual recoil interval corresponds to a range in $\theta_h$ or $\theta_c$ of $\pi/2$, and that the range of $\theta_h$ or $\theta_c$ over $\pi$ encompasses two recoil ranges. In addition, note that, for $v = 0$, we also have $t_a = 0$ and $\theta_c = 0$; while for $v = c$, the ratio $t_a/t_{tap} = \pi/2$, and we also have $\theta_c = \pi/2$.

**Representation of composite recoils along the x-axis: relationship between relative orientation angle and distance along x-axis**

Next let us examine behavior in a localized region within an interimpact interval along the x-axis.

We have seen that for bidirectional motion, the average duration of time after impact, as averaged over many moving recoils, will be proportional to the geometrical complementary half angle between the combining individual recoils that form a composite recoil. Specific values of this angle also correspond to the beginning and end of a full individual simple recoil interval, and they form the limits for composite recoils, and this angle is defined at all intermediate values of composite recoil velocity.

However, this particular relationship between average recoil time duration and angle would be observable only under circumstances such that superpositions of composite recoils occur with all recoil intervals starting at impact times or positions.

Such aligned recurrence together with superposition may take place in the spatial representation of bidirectional motion. Thus, this relationship between average recoil time intervals and angles could become observable in its spatial manifestation along the x-axis. This could show up as a relationship between relative orientation angle and distance along the x-axis.

We will implement this approach by examining the mapping of average recoil time intervals within inter-impact intervals along the x-axis. Impact locations along the x-axis must coincide so as to coordinate the direct Lorentz transformation mappings and the average temporal duration mappings for composite recoils.

Associated with every individual interimpact region along the x-axis, there can be a local mapping of the average time duration of superposed composite recoils onto the x-axis. At every impact position associated with the overall wavelength along the x-axis there can be collocated the mapping of an impact position for composite impacts corresponding to a composite velocity $v = 0$ and a relative orientation or correlation angle $\theta_c = 0$ (modulo $\pi$). Average or apparent time intervals may be mapped along forward and backward directions from each such position, extending to values of the composite velocity $v = c$ and relative orientation or correlation angle $\theta_c = +/- \pi/2$ (modulo $\pi$). Thus, there can be forward and backward mappings of distances corresponding to average time duration or the associated relative orientation or correlation angle that will extend along the x-axis. The mapping is scaled so that the mapped segments extend from the central impact position to halfway to adjacent impact positions, and these segments will adjoin segment mappings associated with adjacent impacts at locations corresponding to impact positions at values of composite recoil speed $v = 0$, and also meeting midway in between impact locations at positions corresponding to composite recoil speed $v = c$. Thus, the total mapping for to- and fro- recoils for each impact position extends over a distance L, or half the wavelength for the bidirectional motion.

In this recoil mapping onto the x-axis, recoils from speed $v = 0$ to speed $v = c$ would extend from a location $x = 0$ at an impact location to a distance $L/2$. Accordingly, in the mapping we would find, as measured from an impact location,

$$v(x)/c = x/(L/2) \qquad (30)$$

Within this mapping, recoil speeds would be in a one-to-one correspondence with recoil distances as mapped along the x-axis over such a range.

Since, at an impact location, the associated relative orientation angle would be $\theta_c = 0$, while at $v = c$, the associated angle would be $\theta_c = \pi/2$, we can see that a single recoil mapping extends over a quarter of the full range of $\theta_c$, and the direct angular mapping over a single recoil along the forward direction would amount to:

$$\theta_c(x) = (\pi/2)[x/(L/2)] = \pi x/L \qquad (31)$$

We can reexpress this as:

$$\theta_c(x) = 2\pi x/\lambda \qquad (32)$$

Here the wavelength characterizing this bidirectional motion is given by twice the recoil mapping distance L, or $\lambda = 2L$.

Alternatively, we can reexpress this in terms of a wave number k, where the wave number is the inverse of the wavelength, and thus k = 1/λ. Thus, we have:

$$\theta_c(x) = 2\pi k x \qquad (33)$$

Furthermore, this angular relationship can be extended indefinitely along the x-axis.

In this interpretation, we can see that for one full cycle of the impact frequency, or one full rotation of the relative orientation angle (the full opening angle going through 2π radians, or $\theta_c$ going through π radians), we only go through half the effective wavelength. This particular effect can be seen to exhibit some resemblance to the characteristic behavior of spin ½ particles, requiring two rotations to complete one full wavelength.

**Relative number of ways that contributing configurations can combine as a function of opening angle for bidirectional motion**

Another aspect of the formation of composite velocity states that is of considerable interest is that it permits a geometrical approach to an evaluation of the relative number of ways that contributing configurations of recoil velocity vectors can combine to form a particular composite state characterized by a specific opening angle between the participating simple recoil motions.

Let us consider geometrically how a composite recoil velocity vector can be formed. We have seen that a composite recoil velocity vector will be directed midway between its two component recoil vectors. These two particular component vectors can form this state, but geometrically so can all other combinations of two such vectors also at half opening angles $\theta_h$ to the composite velocity vector direction, but pointing individually along all of the other directions along a conical surface in three-dimensional space. From these geometrical considerations, it would appear that the number of separate ways that component recoil velocity vectors could contribute to such a configuration would be proportional to the length of the perimeter of the ring of end points of spatial vectors extending to form a cone from a common origin that can contribute to form this net composite velocity state. Since the ring or perimeter of the circle is proportional to $\sin \theta_h$, this indicates that the number of component velocity configurations that can lead to this net composite velocity configuration should be proportional to $\sin \theta_h$. Thus it appears that we may conclude that the relative frequency of states associated with a particular half opening angle $\theta_h$ would be proportional to $\sin \theta_h$.

Qualitatively, we can see from this result that since many more recoil vector combinations would contribute to composite configurations with zero or near zero recoil velocity than to composite configurations recoiling at or near the maximum speed near c, that in the representation along the x-axis we can thus expect many more composite configurations to form with end points at or near impact locations, and fewer composite configurations to form in association with spatial locations along the x-axis that are farther away from impact locations.

**Developing a state function to describe bidirectional to- and fro- motion**

While the submicroscopic impacts and recoils characterizing a particle undergoing vacuum impacts may be regarded as discrete phenomena from a macroscopic point of view, it can be advantageous to adopt continuous functions for their description, so that we can use customary techniques of differentiation and differential equations in the description of the phenomena. We will therefore continue to attempt to develop a functional description to characterize the behavior associated with vacuum impacts and recoils as observed from relatively moving frames of reference.

We can combine the results that we have developed so far into a function describing some major characteristics of macroscopic bidirectional particle motion. As discussed in the preceding section, the relative number of ways that contributing configurations can combine to form a half opening angle $\theta_h$ as a function of $\theta_h$ can be described by the function:

$$\psi(\theta_h) = A \sin \theta_h \tag{34}$$

Here A is an arbitrary multiplicative factor, since this function simply describes the relative frequency of contributing configurations as a function of the angle $\theta_h$. Reexpressing Eqn. (34) in terms of the complementary angle $\theta_c$, we find:

$$\psi(\theta_c) = A \cos \theta_c \tag{35}$$

But the angle $\theta_c$ is also mapped along the x axis, and, in accordance with Eqn. (33), we can express a state function describing bidirectional motion as:

$$\psi(x) = A \cos(2\pi kx) \tag{36}$$

Hence it appears that a particle engaged in macroscopic bidirectional motion can be described by a state function that varies along the x-axis with a sinusoidal amplitude and a wave number dependent on the impact frequency and translational velocity.

**Separating out the contributing constant velocity state functions**

A superposed state of bidirectional to- and fro- constant speed motion has resulted from combining constant velocity free particle motion in each direction along the x-axis. It would be desirable to be able to separate the state function that we have developed for constant speed bidirectional motion into component functions describing constant velocity motions along each direction.

Since a sine function can be expressed in the form of a combination of two exponential functions, we can reexpress the state function describing bidirectional motion in Eqn. (36) as follows:[11]

$$\psi(x) = A \cos(2\pi kx) = (A/2)\{\exp[i2\pi kx] + \exp[-i2\pi kx]\} \tag{37}$$

Thus, we can write the state function describing a bidirectional motion as the sum of two component functions, associated with the forward and backward directions along the x-axis:

$$\psi(x) = \psi_+(x) + \psi_-(x) \tag{38}$$

where:

$$\psi_+(x) = (A/2)\exp(2\pi ikx) \tag{39}$$

and

$$\psi_-(x) = (-A/2)\exp(-2\pi ikx) \tag{40}$$

Eqns. (39) and (40) describe the spatial behavior of functions that would appear to describe the component functions associated with opposite directions of motion.

We have already seen that a traveling wave introduced by Lorentz transformation depends on space and time together in the combination $(t - Vx/c^2)$, if it is traveling to the right. We can incorporate this time dependence into Eqn. (39) by rewriting it as:

$$\psi_+ (x,t) = (A/2) \exp[(2\pi i)(kx-ft)] \qquad (41)$$

and:

$$\psi_- (x,t) = (-A/2) \exp[(2\pi i)(-kx-ft)] \qquad (42)$$

where $k = 1/\lambda = 1/2L$ and $f = f_i/2$.

Thus we have arrived at traveling waves by including a time dependence based upon a common frequency.

    These functions are similar to the free particle waves discussed earlier, and can serve as realizations of the free particle waves discussed in an earlier section of this paper, where we show that, in accordance with the Lorentz transformations in the non-relativistic limit of $\gamma = 1$, the traveling wave associated with a free particle will depend on $(t - Vx/c^2)$.

## 6. THE MORE GENERAL CASE OF A PARTICLE IN NON-UNIFORM MOTION

**What can special relativity tell us?**

    How can we extend analysis of the free particle case so as to obtain some understanding of the situation in which a particle is subject to forces? For a particle that is undergoing motion more complicated than uniform motion with respect to an observer, the special theory of relativity is not strictly applicable and thus cannot give us a definitive description of associated spatial periodicities, since the special theory applies strictly only to the relationships between coordinate systems moving at constant relative velocities. However, consideration of the special theory of relativity can provide us with some insight into possible spatial periodicities and also some related guidance as to the nature of an associated field function. First, for a case of varying velocity we might expect that locally the wavelength would correspond at least approximately to the wavelength expected from special relativity for motion with the local translational velocity. Thus, longer spatial wavelengths might be expected in regions characterized by low translational velocity, and shorter spatial wavelengths might be expected in regions characterized by a higher local translational velocity.

    What can we say about a state function for a more general case? We can anticipate that any such function would also be characterized by impact-related frequencies, and also by the associated wavelengths originating from the relative motion of the frame of reference of the particle with respect to the observer. As seen in the observer's frame, a more general state function to describe impacts and associated recoils might be expected to have to incorporate parameters including impact frequency, wavelength, and magnitude and direction of the macroscopic velocity. We would need to treat it as a function of both space and time in the observer's frame of reference. Generalizing from our earlier work, we might expect that the impact rate or frequency would depend on the total mass, or, since mass and energy are related proportionally, to the associated energy of the particle, in accordance with Equation (2) and Equation (3). The simplest approach to analysis would suggest that locally the wavelength, or the associated curvature of the function, would depend on the local speed associated with the particle's macroscopic translational motion (or the speed associated with bidirectional translational motion) relative to the observer, while the local frequency might depend on the local impact frequency.

**A governing equation for the more general case**

    How would such a state function behave in space and time? The local frequency would presumably be a determinant of the time rate of change of such a state function. As just noted, the most straightforward approach would be to postulate that the local frequency would depend upon the local impact frequency. We know that the impact frequency is proportional to the

particle's mass, in accordance with Eqn. (2). Thus, the time rate-of-change of the state function would also be expected to be proportional to the particle's mass. The magnitude of change in the state function might also be expected to depend on the magnitude of the function itself, if the mechanism of change is based on a fractional effect on the function as a whole. This may also be expected, since we have seen in an earlier section dealing with the case of bidirectional motion that state function amplitudes enter only as relative values rather than as absolute quantities.

Accordingly, we can tentatively write for the time-rate-of-change of the state function:

$$d\psi/dt = (k_1/b)m\psi \qquad (43)$$

Here, $\psi$ is the function describing the particle, m is the particle mass, and $k_1$ is a constant as yet to be determined. Using Einstein's mass-energy relationship, we can reexpress this as:

$$d\psi/dt = (k_1/bc^2)E\psi \qquad (44)$$

This equation is essentially identical in form to the simplest expression of the time-dependent Schroedinger wave equation expressed in terms of the energy or Hamiltonian function.[12]

In Eqn. (44), we can expand the total energy E as a sum of rest energy $E_o$, kinetic energy T, and potential energy U as follows:

$$d\psi/dt = (k_1/bc^2)(E_o + T + U)\psi \qquad (45)$$

The preceding equations (43), (44), and (45) have been based on the vacuum flux impact model. Let us now introduce the classical wave equation that applies generally to a wave that propagates at speed u:

$$\partial^2\psi/\partial x^2 - (1/u^2)\, \partial^2\psi/\partial t^2 = 0 \qquad (46)$$

As discussed earlier, the waves that result from temporal periodicity coupled with the Lorentz transformations propagate at speed $c^2/V$, where V is the speed associated with the relative velocity between the frames of reference of the particle and the observer. Hence these waves would be governed by an equation:

$$\partial^2\psi/\partial x^2 - (V/c^2)^2\, \partial^2\psi/\partial t^2 = 0 \qquad (47)$$

Let evaluate $\partial^2\psi/\partial t^2$ by differentiating Eqn. (43) with respect to the time:

$$\partial^2\psi/\partial t^2 = (k_1/b)\,(\partial m/\partial t)\psi \;+\; (k_1/b)m\,\partial\psi/\partial t \qquad (48)$$

If we limit our attention for the time being to situations in which the mass remains constant and does not change with time, then the first term on the right hand side of the equation will go to zero and we will be left with:

$$\partial^2\psi/\partial t^2 = (k_1/b)m\,\partial\psi/\partial t \qquad (49)$$

Inserting Eqn. (49) into Eqn. (47), we find:

$$\partial^2\psi/\partial x^2 - (V/c^2)^2(k_1/b)m\,\partial\psi/\partial t = 0 \qquad (50)$$

Next, we substitute Eqn. (43) for $d\psi/dt$ in Eqn. (50), and obtain:

$$\partial^2\psi/\partial x^2 - (V/c^2)^2 (k_1/b)^2 m^2 \psi = 0 \tag{51}$$

which can be reexpressed as follows:

$$\partial^2\psi/\partial x^2 - (2mk_1^2/b^2c^4) (mV^2/2)\psi = 0 \tag{52}$$

From Eqn. (52) we can obtain a relationship:

$$(2mk_1^2/b^2c^4)^{-1} \partial^2\psi/\partial x^2 = (mV^2/2)\psi \tag{53}$$

Hence we can obtain the relationship:

$$T\psi = (b^2c^4/2mk_1^2) \partial^2\psi/\partial x^2 \tag{54}$$

where T designates the kinetic energy.
    Let us insert Eqn. (54) into Eqn. (45). We obtain:

$$(bc^2/k_1) \partial\psi/\partial t = (b^2c^4/2mk_1^2) \partial^2\psi/\partial x^2 + E_o \psi + U\psi \tag{55}$$

The resemblance of this equation governing a state function describing vacuum impacts and recoils to the Schroedinger equation is easily recognized. The relationship between the two equations can be made explicit. If we omit the rest energy term and set the quantity $(bc^2/k_1)$ equal to $(ih/2\pi)$, this becomes the time-dependent Schroedinger equation. In fact, the constants b and $k_1$ can even be evaluated separately. Comparing Eqn. (3) with Einstein's energy frequency relationship $E = hf$, we find that $bc^2 = h$. Accordingly, the constant $k_1$ must equal $-2\pi i$.

    Thus, with an appropriate selection of values for the constant coefficients, the governing equation that we have developed for a state function from the vacuum flux impact model turns out to have just the form of the Schroedinger equation.

    The more general equation for the state function describing impact occurrences would thus become:

$$(ih/2\pi) \partial\psi/\partial t = - (h^2/8\pi^2 m) \partial^2\psi/\partial x^2 + U(x,t) \psi \tag{56}$$

This can be recognized as the Schroedinger equation.[12] However, it is notable that this equation has not been developed as a direct description of a particle, but rather it is instead an equation that has been developed to describe the behavior of a function describing the impacts and associated recoils of a particle as seen from a relatively moving frame of reference.

    While this is a very freestyle way of developing a governing equation, it may be of interest in clarifying how the vacuum flux impact model can contribute to providing an interpretation of the Schroedinger equation.

## 7. FURTHER DISCUSSION

    Within the familiar macroscopic world, the when and where of events seem intuitive and straightforward; however, when special relativity enters the picture, less familiar results can ensue. A consequence of special relativity, that an event occurring at a particular location and time in a particular frame of reference is seen to occur at different times and at different locations in any other frame of reference that is in motion relative to the first, has an especially important role in understanding quantum phenomena in this model. This is particularly vivid for the case of a temporal periodicity associated with a particle that repeats throughout all time. As a consequence of the Lorentz transformation, such a temporal periodicity is mapped as both temporal and spatial periodicities in a relatively moving frame of reference associated with an

observer. If each temporal event in the particle's frame of reference corresponds to an impact that has an associated recoil or recoil time period, these may be mapped onto the observer's relatively moving frame of reference. Over each spatial region in the observer's frame of reference the recoil times associated with a large number of recoils can be mapped. It can be seen that in this case special relativity gives us not just a single representation of reality, but multiple superposed representations of reality. It seems possible to conclude at least tentatively that quantum mechanics may be largely about seeing repetitive events from different frames of reference.

In using special relativity to elucidate macroscopic events, it is usual to synchronize clocks and characterize the description as one in a single frame of reference, seen, ideally, by an observer located at a single spatial location, while infinitely repetitive events are not of such great interest. Under those ordinary macroscopic circumstances, superposition does not play an important role. However, in the present problem we are dealing with a macroscopic view of submicroscopic phenomena, and superposition does play an important role. While for macroscopic circumstances, synchronization of clocks throughout the observer's domain is a realistic approach, when dealing with microscopic phenomena occurring very rapidly at very small time intervals, deliberate synchronization of clocks to this accuracy is unrealistic, as macroscopic synchronization of clocks and macroscopic selection of origin of coordinates cannot in principle embody the precision in principle for microscopic events. Thus, while there may be no doubt that temporal and spatial periodicities are present, and we can select a particular function to describe the behavior, its "initial conditions" of coordinate origins appear to be beyond our capabilities of specification. This, together with the overlay of special relativistic observations from a large number of microscopic positions and times in any macroscopic observation would seem to pertain to some aspects of quantum phenomena. The superposition inherent in a fully realized special relativistic state lasting throughout time and extending throughout space may contribute to the ensemble aspect of quantum mechanics.

We find that at least some of the so-called 'wave aspects' of a particle seem to be nothing more than a natural unavoidable concomitant of its existence in a time-average rest frame when it is observed from a relatively moving frame of reference. More of a 'mystery' perhaps is why a single particle can be associated simultaneously with multiple macroscopic frames of reference. In ordinary life, we are accustomed to a one object being intrinsically associated with one frame of reference, specifically its rest frame, and we are uncomfortable with an observer simultaneously seeing one object in multiple frames of reference. But so-called quantum objects – that is to say, objects – may routinely exist in multiple frames of reference; while what we recognize as classical are objects associated with a single frame of reference. Perhaps what we think to call a classical object seems to require a single frame of reference and also a single location, or for extended objects, a contiguous extended location.

The fact that a particle can appear to be present in multiple locations simultaneously from the point of view of a relatively moving observer can also lead to issues relating to causality, and to the speed of information transfer, and the characterization of the locality of the particle.

Because the vacuum flux impacting a particle can come from all directions in three dimensional space, from geometrical considerations of solid angle it would appear that when a recoiling particle is struck by the next impact, it will on average tend to move in a direction perpendicular to its initial recoil direction. Thus, succeeding recoils will tend to be perpendicular to each other statistically. This may be related to the need for complex state functions and governing equation, and to the circumstance that the time rate of change of the state function turns out to be imaginary with respect to the Hamiltonian operator applied to the state function.

Because of the perpendicularity of succeeding recoils, there may be some reason to consider also a frequency that is half the impact frequency, corresponding to two impacts per cycle; this is illustrative of the possible importance of harmonics or subharmonics that could indicate structure in sequences of impacts. More generally, we can consider the possibility of other temporal structures superimposed on the basic Lorentz transformation structure resulting from the original impact frequency.

Additional quantum-like effects that could a also be a consequence of the stochastic behavior of matter under vacuum flux impacts might include particle spin, which perhaps might be interpretable as a characteristic associated with the perception in a relatively moving frame of reference of the particle's stochastic motion under vacuum flux bombardment rather than as an intrinsic characteristic of the particle itself.

It is notable that this model also connects fundamentally local properties to the fundamentally global approach of quantum mechanics. Also, the connection developed in this model between gravitation and quantum phenomena indicates that the nature of the physical vacuum affects the properties of space-time and the properties of particles as well.

## 8. SUMMARY/CONCLUDING REMARKS

This paper is intended as an attempt to gain some fresh insight into and a further degree of understanding of some of the counterintuitive aspects of quantum mechanics.

The vacuum flux impact model proposes a specific physical mechanism to account for the probabilistic nature of quantum phenomena. In this model, quantum phenomena emerge as a description of fluctuations around an average in which classical gravitational physics on a spatial background emerges from the basic theory. Special relativity plays an essential role in relating quasiperiodic temporal phenomena in different relatively moving coordinate systems.

The vacuum flux impact model interprets physical phenomena as originating from vacuum flux impacts on particles of matter. We develop a representation in the observer's frame of reference of the repeated impacts in the particle's frame of reference. While we may hopefully infer some information about the behavior of the particle itself, since the presence of a particle is necessary for the occurrence of impacts and recoils, this representation is primarily about impact events. We use this information to describe the behavior of impacts and their associated recoils rather than to attempt to describe the particle directly as such.

These repeated flux impacts on any matter particle define a frequency associated with that particle that is proportional to the particle mass. Since a particle of matter is characterized by an impact frequency in its time-averaged macroscopic rest frame, it will appear when viewed from a relatively moving frame to be characterized by a spatial as well as a temporal periodicity, in accordance with special relativity. The wave motion associated with this spatio-temporal periodicity exhibits frequencies proportional to the energy and wavelengths inversely proportional to the momentum.

Analysis of superposed motion leads to the development of state functions describing bidirectional free particle motion as well as state functions describing constant velocity motion.

Further examination leads to the development of a governing equation resembling the Schroedinger equation to describe the behavior of particle impacts and recoils in the vacuum flux impact model.

According to this model we interpret quantum phenomena as deriving ultimately from vacuum flux impacts on matter particles and the mapping of impacts and recoils onto an observer's frame of reference in accordance with special relativity. Although elementary quantum mechanics associated with the Schroedinger equation is usually characterized as non-relativistic quantum mechanics in that it describes non-relativistic particle motion, in this analysis it is intrinsically based on special relativity.

It is hoped that these results will provide a useful formulation permitting further insight into the nature of some physical phenomena.

---------------------------------------------------------------------------------------------------------

**Résumé**
La relativité spéciale combineé avec le modèle stochastique d'impact de flux dans le vide mène à une interprétation explicite de plusieurs phénomènes de la mécanique quantique élémentaire. Nous observons des caractéristiques de le particule submicroscopique heurtée fréquemment, en même temps que nous observons la manière dont les effets lies au comportement de le particule apparaissent dans un autre système de référence en mouvement. Vu des systemes de référence en mouvement, le temps et les endroits des impacts et des reculs montrent automatiquement le comportement de l'onde. Ce modèle amene aux ondes associées aux particules ayant des fréquences proportionnelles à l'énergie et des longueurs inversement proportionnelle à la force du mouvement. Si les ondes associeés aux particules se déplaçant dans des directions opposes avec des vitesses constantes sont superposées, on obtient une amplitude qui change sinusoïdalement avec la distance. Observés du relatif système de reference en mouvement, les impacts et les reculs peuvent sembler se produire simultanément en plusieurs endroits dans l'espace. On développe une équation régissante identique dans la forme à l'équation de Schroedinger qui décrit le comportement des impacts et des reculs qui en résultent. Ceci permet d'observer de nombreux aspects de la mécanique quantique d'une manière intuitivement visuelle.

**References**

1. C. L. Herzenberg, Physics Essays **13**, 604 (2000).
2. E. Nelson, Phys. Rev. **150**, 1079 (1966).
3. W. Weitzel, Zeit. für Physik **134**, 264 (1953).
4. L. de Broglie, J. Phys. **5**, 225 (1927).
5. L. de Broglie, *An Introduction to the Study of Wave Mechanics* (Methuen, London, 1930).
6. L. de Broglie, *Théorie générale des particules à spin* (Gauthier-Villars, Paris, 1943).
7. E. P. Ney, *Electromagnetism and Relativity* (Harper and Row, New York, 1962), p. 40.
8. P. G. Bergmann, *Introduction to the Theory of Relativity* (Prentice-Hall, Inc., New York, 1942), p. 40.
9. G. Stephenson and C. W. Kilmister, *Special Relativity for Physicists*, (Longmans, Green & Co., New York, 1958), p. 41-42.
10. R. S. Burington, *Handbook of Mathematical Tables and Formulas*, 3rd edition, (Handbook Publishers, Sandusky, Ohio, 1949), p. 65.
11. R. S. Burington, *Handbook of Mathematical Tables and Formulas*, 3rd edition, (Handbook Publishers, Sandusky, Ohio, 1949), p. 24.
12. L. I. Schiff, *Quantum Mechanics* (McGraw-Hill, New York, 1949), p. 21.

**C. L. Herzenberg**
Herzenberg Associates
1700 E. 56th Street, Suite 2707
Chicago, IL 60637-5092 U.S.A.

e-mail: carol@herzenberg.net